\documentclass[sigplan,nonacm]{acmart}
\usepackage[]{hyperref}
\usepackage{fancyhdr} 

\usepackage{array}
\usepackage{makecell} 

\usepackage{algorithmicx}
\usepackage{algpseudocode}
\usepackage{algorithm}
\usepackage{multirow}
\usepackage{subcaption}
\usepackage{graphicx}
\usepackage{geometry}

\begin{document}

\title{Architectures for Heterogeneous Quantum Error Correction Codes}

\author{Samuel Stein$^{1}$, Shifan Xu$^{2,3}$, Andrew W. Cross$^{5}$, Theodore J. Yoder$^{5}$, Ali Javadi-Abhari$^{5}$, Chenxu Liu$^{1}$, Kun Liu$^{3,4}$, Zeyuan Zhou$^{3,4}$, Charles Guinn$^{6}$, Yufei Ding$^{7}$, Yongshan Ding$^{2,3,4}$, Ang Li$^{1,8}$\\[1em]}

\affiliation{%
    $^{1}$ Pacific Northwest National Laboratory, Richland, WA, USA \\
    $^{2}$ Departments of Applied Physics, Yale University, New Haven, CT, USA \\
    $^{3}$ Yale Quantum Institute, Yale University, New Haven, CT, USA \\ 
    $^{4}$ Department of Computer Science, Yale University, New Haven, CT, USA \\  
    $^{5}$ IBM Quantum, IBM T. J. Watson Research Center, Yorktown Heights, NY, USA \\
    $^{6}$ Department of Physics, Princeton University, Princeton, NJ, USA \\ 
    $^{7}$ Department of Computer Science \& Engineering, University of Califronia San Diego, San Diego, CA, USA \\
    $^{8}$ Department of Electrical \& Computer Engineering, University of Washington, Washington, WA, USA \\[1em]
    \country{}
}

\begin{abstract}

Quantum Error Correction (QEC) is essential for future quantum computers due to its ability to exponentially suppress physical errors. The surface code is a leading error-correcting code candidate because of its local topological structure, experimentally achievable thresholds, and support for universal gate operations with magic states. However, its physical overhead scales quadratically with number of correctable errors. Conversely, quantum low-density parity-check (qLDPC) codes offer superior scaling but lack, on their own, a clear path to universal logical computation. Therefore, it is becoming increasingly evident that there are significant advantages to designing architectures using multiple codes. Heterogeneous architectures provide a clear path to universal logical computation as well as the ability to access different resource trade offs.

To address this, we propose integrating the surface code and gross code using an ancilla bus for inter-code data movement. This approach involves managing trade-offs, including qubit overhead, a constrained instruction set, and gross code (memory) routing and management. While our focus is on the gross-surface code architecture, our method is adaptable to any code combination and the constraints generated by that specific architecture.

Motivated by the potential reduction of physical qubit overhead, an ever important feature in the realization of fault tolerant computation, we perform the first full system study of heterogeneous error-correcting codes, discovering architectural trade-offs and optimizing around them. We demonstrate physical qubit reductions of up to $6.42\times$ when executing an algorithm to a specific logical error rate, at the cost of up to a $3.43\times$ increase in execution time.

\end{abstract}

\maketitle
\pagestyle{fancy}
\fancyhf{} 
\renewcommand{\headrulewidth}{0pt} 

\section{Introduction}

Quantum error correction (QEC) offers a potential solution to the noise problem in quantum systems by exponentially suppressing errors, a key factor in achieving error rates low enough for practical quantum computing. The Surface Code stands out as one of the most studied QEC codes \cite{fowler2012surface}. It requires only local interactions, provides a threshold of $~ 0.7\%$, and can perform universal quantum computation when combined with a magic state factory \cite{fowler2012surface}.

A well studied low overhead approach for computation on fault-tolerant systems involves the Clifford + T formalism, where Clifford gates are executed fault-tolerantly, and T gates are executed via magic state injection \cite{litinski2019magic}. Pauli-Based Computation is built on this premise , enabling significant optimizations by commuting and pruning Clifford gates, while leaving circuit blocks consisting of solely non-Clifford gates and measurements.

The surface code solution, however, comes with expensive, if not prohibitive cost when scaling, as the physical overheads expands quadratically with the number of correctable error. One promising family of codes that are more physically scalable are the quantum Low Density Parity Check (qLDPC) codes, which offer significantly better logical qubit per physical qubit yields \cite{breuckmann2021quantum}. The gross code, a recent qLDPC variant, especially showcases encouraging results, offering surface-code-level phyiscal error rate thresholds and a net logical-to-physical qubit encoding rate of 1/24 \cite{bravyi2024high}.

Nevertheless, universal computation for qLDPC codes like the gross code is not yet at a similar level of maturity as computation in the surface code. For example, while Clifford gates on the gross code can be performed using surgery techniques \cite{cross2024linear}, it is not clear how much time overhead may be incurred in practice. Additionally, while non-Clifford gates on the gross code can be implemented by using magic states, the state-of-art magic state preparation protocols are based on the surface code architecture \cite{litinski2019magic,gidney2024magic}. These factors mean the gross code or other qLDPC codes do not currently serve as one-code solutions to fault-tolerant quantum computation.

\begin{figure}
    \centering
    \includegraphics[width=1\linewidth]{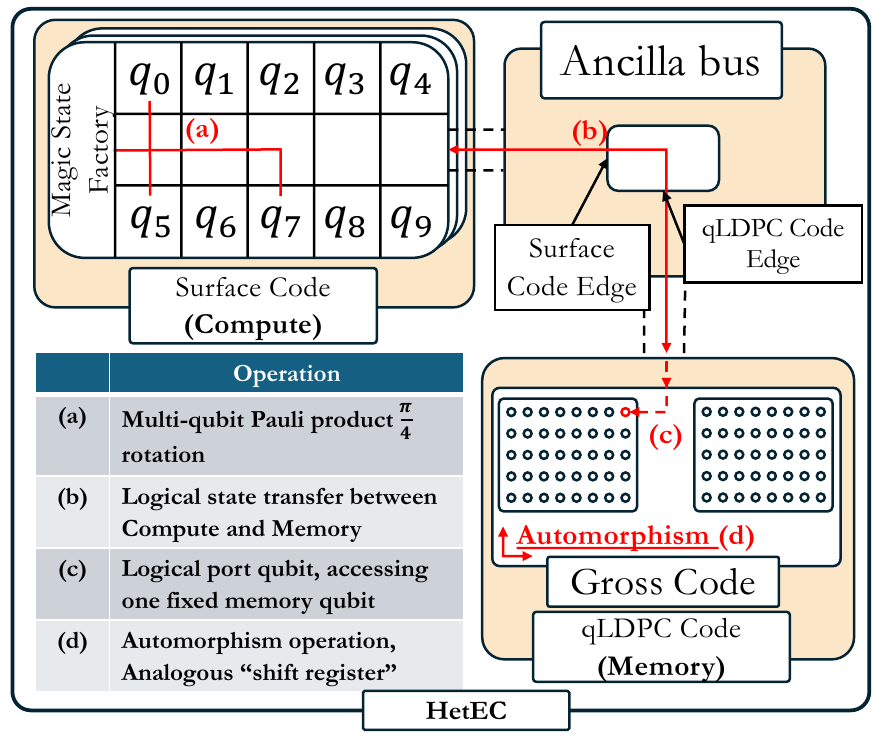}
    \caption{Heterogeneous Error Correcting Architecture sketch. Red underlined text highlights operations required to operate computation between codes, and will be described in the paper.}
    \label{fig:enter-label}
\end{figure}

Here we propose a heterogeneous architecture to take advantage of the strengths of both the surface code and the gross code. Similarly to the von-Neumann paradigm for classical computation, we separate the computer into a processing unit composed of surface code blocks, a memory unit composed of gross code blocks, and an ancilla bus that transfers quantum information between them \cite{cohen2022low,cross2024linear}. This solution will allow the surface-code based quantum CPU to perform universal but expensive non-Clifford and Clifford operations, at an affordable scale; leveraging the gross-code based quantum memory to store and execute low cost Clifford operations at a much larger scale; and in the meanwhile exploiting locality and mitigating quantum memory access' delay.

In this work, we propose HetEC, a surface-gross heterogeneous code architecture that explores the design space of coupling the codes, scheduling code teleportation, and transpiling logical quantum circuits to the hybrid architecture. We tackle unique challenges of performing computation across these codes, discovering and tackling challenges such as non-Clifford Pauli patch measurement weight constraints, asynchronous logical clock cycles, logical data management, movement within the gross code memory and inter-code data movement. We attempt to evaluate and clarify trade-offs in physical qubit requirements, circuit depth, and logical error rates in the development of heterogeneous error-correcting architectures

HetEC makes the following key contributions:

\begin{itemize}
    \item \textbf{Heterogeneous Quantum Error Correction Architecture}: We propose a hybrid quantum error-correction architecture that combines the gross code (memory) and the surface code (compute) via an ancilla bus. We address challenges such as asynchronous logical clock cycles, limitations on Pauli string weights due to the surface code's computational capacity, and data movement within and between the gross code memory and surface code.

    \item \textbf{Transpilation to Heterogeneous Codes}: We develop \textit{HetEC}, a transpiler and scheduler that maps arbitrary gate-based quantum circuits to our hybrid architecture within the Pauli-based computation model, respecting code and system constraints. HetEC effectively manages logical qubit permutations, tackles the challenges of transpiling Pauli-based measurements with limited string weights, and schedules asynchronous operations across the gross code and surface code components.

    \item \textbf{Comprehensive Trade-Off Analysis}: We evaluate the space, time, and error rate trade-offs of the proposed hybrid architecture on three benchmark algorithms. Our analysis shows up to a \textbf{6.42×} reduction in physical qubit requirements at the cost of up to a \textbf{3.43×} increase in logical clock cycles to achieve a targeted logical error rate. These findings offer insights into the practical benefits and trade-offs of hybrid error-correction architectures.
\end{itemize}

\begin{figure*}
    \includegraphics[width=\textwidth]{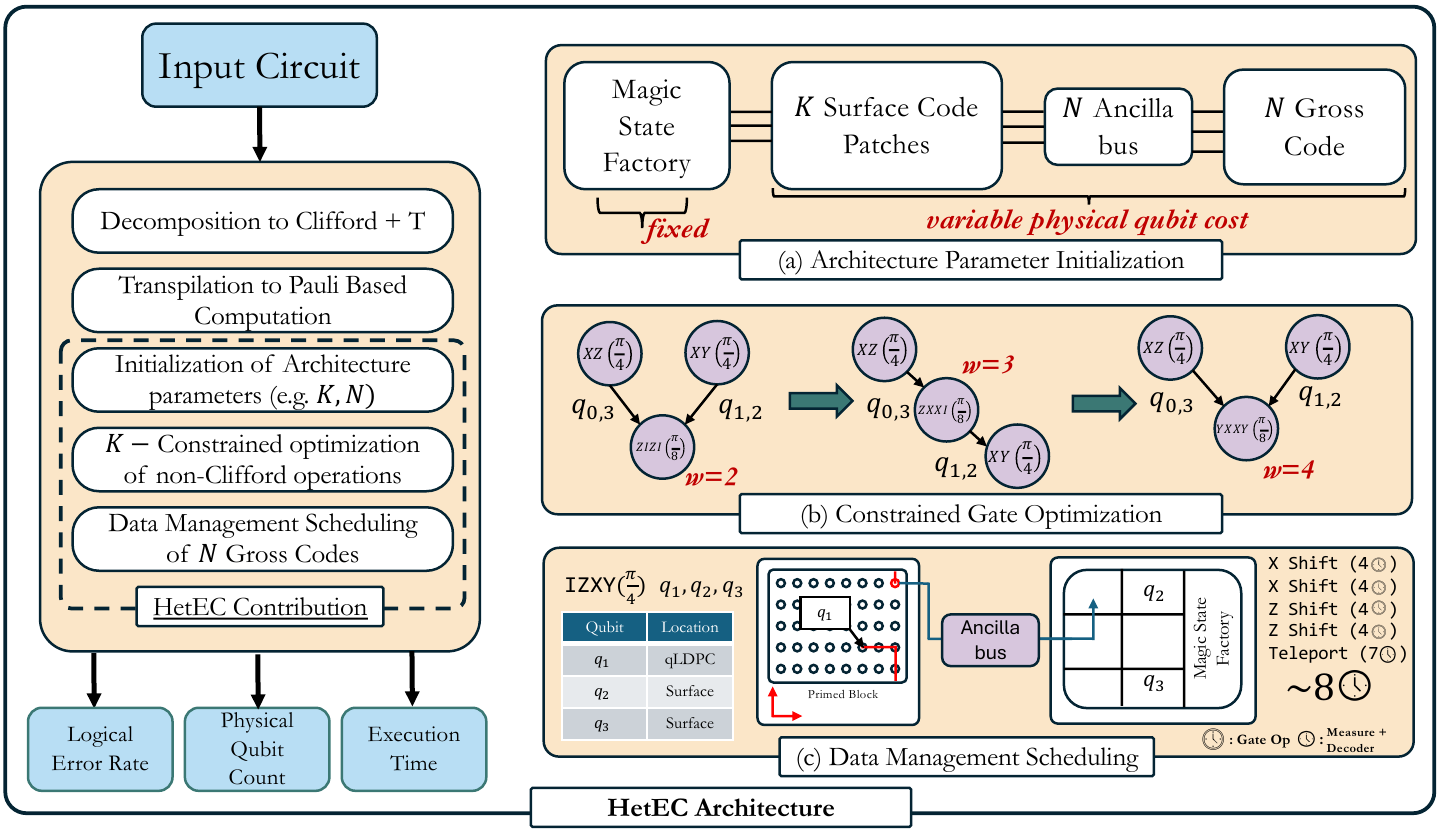}
    \caption{Depiction of HetEC workflow and architecture. Input circuits are passed into HetEC, and transpiled to a user-defined specific combination of gross Code and surface Code tiles. Optimization of Clifford and non-Clifford gates are constrained by the defined setup. Circuits are scheduled over the proposed architecture and logical error probabilities and execution time computed based on the transpiled circuit.}
    \label{fig:arch_fig}
\end{figure*}

\section{Background}
\label{sec:background}
\subsection{Quantum Error Correction}

A quantum error correcting code, encoding k logical qubits into n physical qubits, is a $2^k$ dimensional logical subspace of a larger $2^n$ dimensional Hilbert space. This logical subspace is stabilized by stabilizers $S$, defined by the condition ${S_i|\psi\rangle = |\psi\rangle,\forall {S}}$ \cite{fowler2012surface}. 

When a single error occurs and we measure stabilizer $S_i$ that involves the affected qubit, the stabilizer measures $-1$ parity, indicating an error on one of the qubits. During each round of error correction, all stabilizers are measured, and this information decoded \cite{wu2023fusion}. The decoder determines the likely qubit errors based on the observed measurements.

This process is repeated and interleaved with computations to ensure fault tolerance, detecting and correcting errors that occur between cycles \cite{sayginel2024fault}. If the predicted correction is incorrect, a logical error occurs. Each code has a pseudothreshold, which is the physical error probability below which the logical error rate is less than the lowest physical error rate. Codes are described by their code parameters [[n,k,d]], where $n$ is the number of physical qubits, $k$ is the number of logical qubits, and $d$ is the code distance, which is the minimum number of errors causing an undetectable logical error.

\subsection{Surface Code}

The Surface Code is defined by stabilizers acting on weight-4 plaquettes on a 2D lattice. Stabilizers are placed on a $d \times d$ grid of plaquettes \cite{fowler2012surface}. The Surface Code has parameters $[[d^2, 1, d]]$, scaling quadratically with the number of correctable errors.

Multi-qubit operations in the Surface Code are performed via lattice surgery \cite{litinski2019game}. This is done by introducing new stabilizers between surface code patches \cite{litinski2019game}. This is accomplished over an ancilla surface code path.

All Clifford operators can be accomplished using lattice surgery and transversal gates such as the Hadamard gate \cite{fowler2012surface,litinski2019magic}, and non-Clifford operations, such as the T gate, are performed using magic state injection \cite{litinski2019magic}. Together, these operations provide a universal gate set.

\subsection{qLDPC Codes and The Gross Code}

Quantum Low Density Parity Check (qLDPC) code are a family of codes that are characterized by a sparse parity check matrix. Surface codes are within this family, though we use the term qLDPC codes to refer to high-rate codes \cite{bravyi2024high}. The parameters of qLDPC codes vary depending on the specific code family, such as Hypergraph Product codes \cite{tillich2013quantum,xu2024constant} or bivariate bicycle codes \cite{bravyi2024high}. 

In this paper, we focus on the qLDPC code known as the gross code. The gross code demonstrates a high error threshold of 0.65\% $p_{physical}$, while significantly reducing physical qubit overhead when compared to the surface code \cite{bravyi2024high}. The gross code is notable for its low-depth syndrome measurement circuits, efficient two-layer hardware implementation, and readily accessible logical operators, making it a promising candidate for fault-tolerant computation. The code supports Clifford operations on 11 of the 12 logical qubits \cite{cross2024linear,bravyi2024high}. In this section, we highlight the key architectural considerations when utilizing the gross code, but refer the reader to \cite{bravyi2024high} for a detailed description of its construction and logical operations.

\begin{figure*}
    \includegraphics[width=1\textwidth]{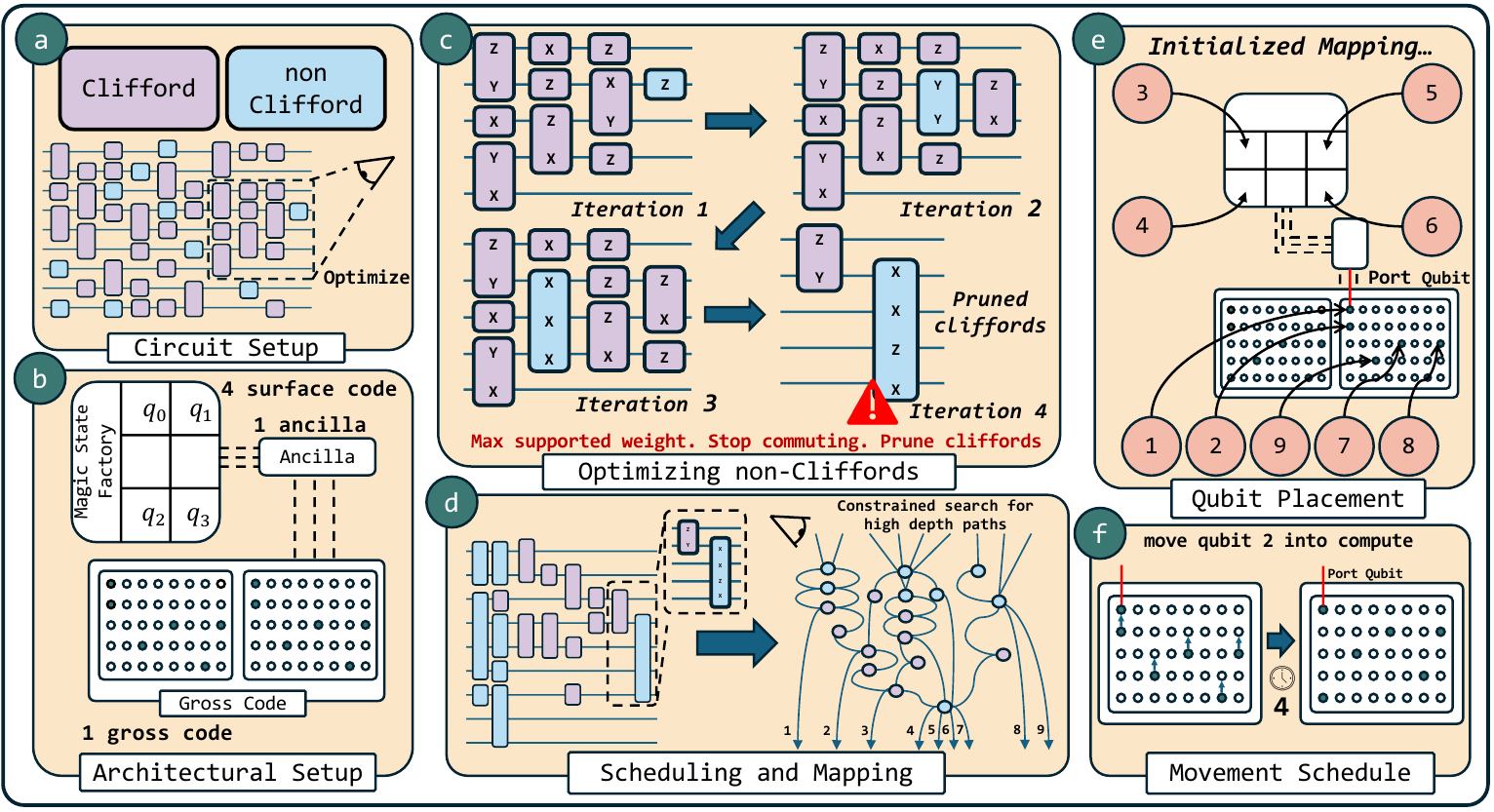}
    \caption{Depiction of the HetEC workflow, compiling and scheduling a circuit for a heterogeneous architecture. The workflow involves (a) loading a circuit in the Pauli-based computational model for optimization, constrained by the architecture defined in (b), where the gross code structure and surface code count are set. (c) Gates are optimized according to these architectural constraints, resulting in a shorter circuit with larger Pauli product measurements. (d) The new circuit is traversed to identify maximum depth gate sets supported by the surface codes, while gates requiring gross code (memory) access are scheduled to maximize gate support and minimize data movement. (e) Initial qubit mappings are determined by the gate sequence, with qubits placed in memory according to pending movements. (f) Automorphisms are applied to permute logical qubits in the memory block, granting access to the port qubit and enabling data movement between the surface code and gross code (memory). }
    \label{fig:compilation_procedure}
\end{figure*}

\subsubsection{Code Structure and Automorphism Gates}

The gross code is a [[144,12,12]] code from the Bivariate Bicycle (BB) code family, which are CSS codes. As it encodes multiple logical qubits, the ability to individually address each logical qubit with low overhead is crucial in qLDPC code design. \emph{Automorphism gates} provide the ability to implement fault-tolerant in-block permutations of logical qubits. Thus, the gross code can access all the logical qubits with only one ``probe'' logical qubit attached to the ancillae system, as permuting the ``probe'' qubit with other logical qubits using automorphism gates generates the entire logical qubit set.

Specifically, the automorphism gates rely only on the connectivity already claimed by the parity check matrices, without introducing additional connectivity and physical qubits. They are fault-tolerant in-block gates implementable by low-depth CNOT in parallel. We elaborate more details of the automorphism gates and their function on logical qubit movement for HetEC architecture in Section\ref{sec:hetec}.







\subsection{Transpilation to Pauli Based Computation}

When transpiling an arbitrary quantum circuit into the Pauli-based computation formalism, we first decompose it into the Clifford+T gate set, comprising $H$, $S$, $T$, and $\text{CX}$ gates \cite{bravyi2005universal}. These gates are transformed into sequences of Pauli rotations, represented as $e^{i \theta P}$, where $P \in \{I, X, Y, Z\}$ and $\theta$ is the rotation angle. 
The standard decompositions are: $H = Z\left(\tfrac{\pi}{4}\right) X\left(\tfrac{\pi}{4}\right) Z\left(\tfrac{\pi}{4}\right)$, $S = Z\left(\tfrac{\pi}{4}\right)$, $T = Z\left(\tfrac{\pi}{8}\right)$, and $\text{CX} = \text{ZX}\left(\tfrac{\pi}{4}\right) \text{ZI}\left(-\tfrac{\pi}{4}\right) \text{IX}\left(-\tfrac{\pi}{4}\right)$.

With these transformations, we can optimize non-Clifford gates by commuting them behind Clifford gates. We can commute the Clifford Pauli product rotations ($\theta = \frac{\pi}{4},P )$ past the non-Clifford Pauli product rotations ($\theta = \frac{\pi}{8},P'$) as follows: if \( PP' = P'P \), i.e. the operators commute, we can move $P$ past $P'$ trivially without any change in operator. If they anti commute, i.e. \( PP' = -P'P \), moving $P$ past $P'$ will introduce a phase factor to $P'$, transforming $P'$ into $iPP'$ for the non-clifford operator.

Finally, we optimize measurement operations by absorbing Clifford gates into the measurement operators. This absorption allows all Clifford gates to be effectively removed from the circuit, simplifying the overall computation. This removal of Cliffrd gates however is not necessarily possible in the heterogeneous setting, and will be discussed under Section \ref{sec:hetec}.

\subsubsection{BB \& Surface Code Ancilla bus Construction}

The fault tolerant movement of logical information between codes is an important technique in the quantum error correction landscape, and allows for the state transfer of some logical state between two codes,  transforming $|\Psi\rangle_{L(A)}|0\rangle_{L(B)}$ into $|0\rangle_{L(A)}|\Psi\rangle_{L(B)}$. This can be accomplished by utilizing a specially constructed ancilla system (here, referred to as an ancillary bus) and recent qLDPC surgery techniques. Originally, the ancillary bus was conceived as a hypergraph product of a repetition code and a Tanner subgraph of the qLDPC code \cite{cohen2022low}, which resulted in an $\Theta(d^2)$ size system. Recent advances \cite{cross2024linear, williamson2024low,ide2024fault,swaroop2024universal} have improved the ancillary bus size to near linear in $d$ (with log factors in worst cases). Most relevant to the present work, Ref.~\cite{cross2024linear} assembled a 103 qubit ancillary bus that enables the implementation of the full Clifford group on 11 out of the 12 logical qubits in the gross code. To enable information transfer between the gross code and a surface code, this 103 qubit ancillary bus can also be connected to the surface code using the LDPC adapter idea \cite{swaroop2024universal} without additional qubit overhead. The HetEC framework relies heavily on this 103 qubit ancillary bus to create a heterogeneous computer. 

Notably, we acknowledge the existence of theoretical foundations for code teleportation. Our work is focused on designing the architecture and layout for coupling the surface-gross codes, the scheduler for managing logical data movement and scheduling operations within the heterogeneous architecture, and the entire compilation and transpilation process to these heterogeneous architectures.



\section{HetEC Architecture}
\label{sec:hetec}



In this section, we propose HetEC, a heterogeneous error correction architecture that combines the computational advantages of the surface code with the memory-like properties of the gross code.

We optimize to maximize the benefits of each code, using the surface Code for both Clifford and non-Clifford operations and the gross code for logical memory with a supported Clifford gate set, whilst optimizing towards minimizing the cost of operating heterogeneously. 

Operating heterogeneously introduces several new challenges, which we discuss in detail below. Section 3.1 describes an architectural overview of HetEC, 3.2 the logical data management within the gross code and between codes, 3.3 the transpilation to HetEC, and 3.4 the asynchronous logical clock cycles between codes and how to adapt operation scheduling accordingly.

\subsection{HetEC Overall Architecture}

\begin{table*}[t]
    \centering
    \resizebox{\textwidth}{!}{ 
    \begin{tabular}{ p{3.2cm}||p{2cm}|p{3cm}|p{3cm}|p{3cm}|p{3cm}}
     \hline
     \hline
     Instructions & Ancillary Qubit Count & Logical Clock Speed & Logical Error Rate under $p=10^{-3}$ & Logical Error Rate under $p=10^{-4}$ & Logical Error Rate under $p=10^{-5}$ \\
     \hline
     Clifford Pauli Rotation & 103 & 14 \cite{cross2024linear} & $4 \times 10^{-5}$ & $2\times10^{-9}$ & $ 6\times10^{-14}$ \\
     \hline
     Logical Qubit Automorphism  & N/A & 1 \cite{bravyi2024high}  & $4 \times 10^{-7}$ & $4 \times 10^{-12}$ & $4 \times 10^{-17}$ \\
     \hline
     X/Z Measurement & 103 & 7 \cite{cross2024linear} & $2 \times 10^{-5}$ & $8\times10^{-10}$ & $ 3\times10^{-14}$ \\
    \hline
     Joint XX Measurement & 103 & 7 \cite{cross2024linear}& $2 \times 10^{-5}$ & $8\times10^{-10}$ & $ 3\times10^{-14}$ \\
    \hline 
    \end{tabular}
    } 
    \caption{Instructions in gross code with simulated and estimated error probability, assuming the same physical gate error $p$ for both single-qubit Pauli and two-qubit CNOT gates. Logical error rates under physical error rates $p=10^{-3}$ and $p=10^{-4}$ are from simulation results of \cite{cross2024linear,bravyi2024high,xu2024constant}; logical error rate under physical error rate $p=10^{-5}$ is fitted from \cite{cross2024linear,bravyi2024high}.}
    \label{tab:QLDPCinst}
\end{table*}

\begin{table*}[t]
    \small
    \centering
    \resizebox{\textwidth}{!}{ 
    \begin{tabular}{ p{4cm}||p{3cm}|p{4cm}}
     \hline
     \hline
       Operations & Logical Clock Speed & Logical Error Probability \\
     \hline
     Clifford Gate & $d$ \cite{chamberland2022universal,litinski2018lattice,cross2024linear} & $0.03(\frac{p}{0.01})^\frac{d+1}{2}$ \\
     \hline
     non-Clifford gate & $2d$ \cite{litinski2019game} & $w_{pauli}(0.03(\frac{p}{0.01})^\frac{d+1}{2})$ \\
     \hline
    \end{tabular}
    } 
    \caption{Instructions in a distance-d surface code with estimated error probability from \cite{beverland2022assessing}, assuming physical error rate $p$. $w_{pauli}$ refers to the number of pauli operators in the non-clifford operation. T state error itself is considered negligible, motivated by the recent low-overhead magic state cultivation scheme \cite{gidney2024magic}.}
    \label{tab:surinst}
\end{table*}

Our heterogeneous architecture consists of three components: \textbf{surface code blocks}, \textbf{qLDPC code blocks}, and \textbf{ancillary qubit blocks} (ancilla bus) that facilitate data movement between the codes (see Figure~\ref{fig:compilation_procedure}(b)).

\textbf{Surface code blocks} support both Clifford and non-Clifford computations assisted by magic states. We envision a layout similar to \cite{litinski2019game}, featuring a planar checkerboard of surface code tiles with adjacent ancillas and an appended magic state factory. This setup allows for lattice surgery between arbitrary surface code patches, with practical implementations proposed in superconducting systems~\cite{kosen2022building,bravyi2024high,smith2022scaling,acharya2024quantum}. We assume a physical qubit overhead of $1.5\times$ the number of dedicated surface code compute modules, accounting for ancilla patches used for routing and entangling operations. 

\textbf{Gross code blocks} serve as logical memory, handling qubits not actively involved in non-Clifford operations and executing Clifford gates. We choose the gross code for its research maturity and compatibility with superconducting systems, though any other qLDPC code can be substituted. The gross code interfaces with the ancilla bus via a single logical qubit, and logical qubit automorphisms enable internal qubit routing. By sacrificing one logical qubit, the gross code supports the full set of Clifford operations. 

\textbf{Ancilla buses} provide connectivity between the surface code and gross code blocks, enabling data movement between codes. In our scheme, we couple the ancilla bus to a single "port" qubit on the gross code to minimize physical challenges and overhead associated with coupling to all logical qubits. The gross-surface code ancilla bus consists of 103 physical qubits, enabling bi-directional data movement. 

Table~\ref{tab:QLDPCinst} presents the instruction set, logical error rates, and clock rates for the HetEC architecture. This architecture is characterized by three key features that illustrate the trade-offs in this design space:

\begin{enumerate}
    \item \textbf{Physical Qubit Count}: The total number of physical qubits used in the gross code, surface code, and ancilla bus blocks (excluding magic state factories, which are constant across architectures).

    \item \textbf{Surface-Gross Compute-Memory Overhead}: Since all non-Clifford operations must be performed on the surface code, overhead arises from managing logical qubit placement, scheduling data movement between codes (shorthanded as "IO" - Input/Output), and optimizing operation ordering.

    \item \textbf{Compilation and Transpilation to HetEC}: Operating on a heterogeneous code architecture imposes nuanced constraints on non-Clifford operations—for example, optimizing away Cliffords is not always possible. Data movement operations must be inserted accordingly. Asynchronous logical clock cycles, due to variance in syndrome measurement rounds, complicate scheduling. These challenges contribute to the overhead in an algorithm's depth.
\end{enumerate}

By exploring different architectural designs tailored to specific future quantum algorithms, we can evaluate the trade-offs that are availed to us by operating heterogeneously.

\subsection{HetEC Data Management}

Fully coupling all $12$ of the gross code’s logical qubits to the logical ancilla would impose a significant physical overhead, requiring over 1,200 physical qubits, in addition to other challenges, such as overlaps between qubits supporting different logical operators. As such, we propose to attach the to only one "probe" qubit of the logical qubits in the gross code \cite{cross2024linear}, as seen in Fig \ref{fig:arch_fig}-(c) and \ref{fig:compilation_procedure}-(f). 

\begin{figure*}
    \includegraphics[width=1\textwidth]{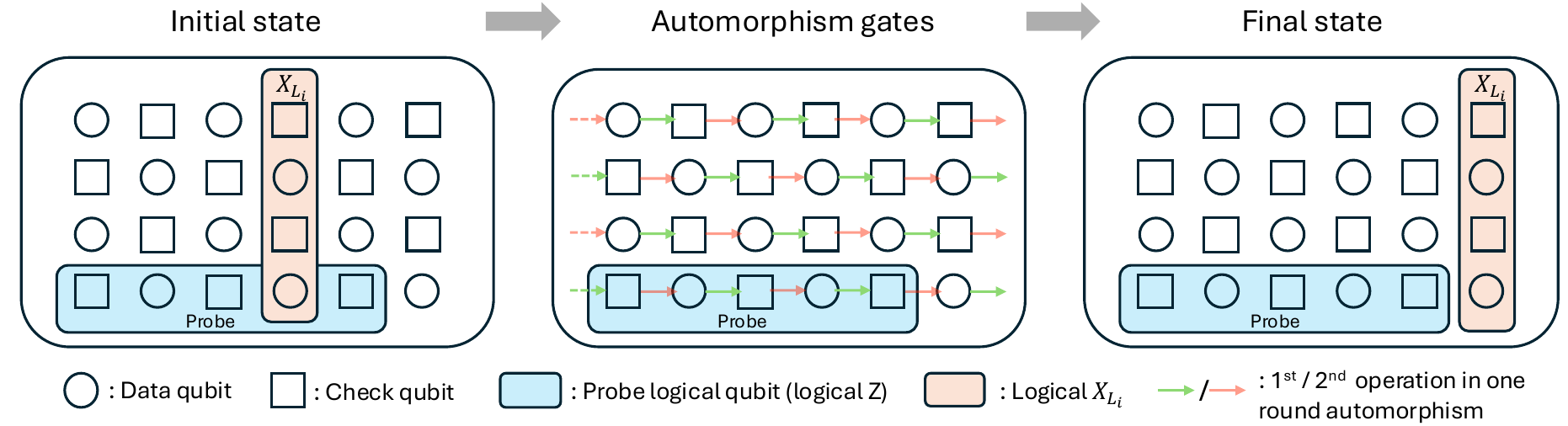}
    \caption{An automorphism consists of two operations: first, swapping data qubits with check qubits (green arrows), and then swapping check qubits with a different set of data qubits (red arrows). These gates are executed in parallel using the connectivity defined by the BB code parity check matrices, with dashed arrows indicating toric boundaries. After the sequence, the physical qubits supporting the logical operator $X_{L_I}$ are relocated, which alters the logical $Z$ operator due to the fixed position of the probe qubits. Since the probe qubits are attached to the ancilla bus and cannot be moved, the logical $Z$ operator changes as a result.}
    \label{fig:automorphism fig}
\end{figure*}

We propose a two-stage qubit movement protocol that combines logical qubit automorphisms with state teleportation to move a target qubit from the gross code to the surface code. First, the automorphism permutes the logical qubits in the gross code block, shifting the target qubit to the port, as illustrated in Fig \ref{fig:automorphism fig} and \ref{fig:compilation_procedure}-(f). In the second stage, state teleportation between the gross and surface code blocks is performed via a joint XX measurement, followed by corrective pauli gates based on the observed measurement. 



When the ancilla is coupled to either the X or Z logical operators from both the surface and gross blocks, the ancilla system enables bi-directional state teleportation, similar to the measurement-based CNOT used in surface code lattice surgery. For the structure of our ancilla system the mono-layer approach is used \cite{cross2024linear}. The mono-layer approach is resource-efficient and tailored specifically for the gross code. 

The execution of gross code automorphisms and state teleportation incurs overhead in terms of both execution time and logical error rate. Hence, we want to minimize the costs associated with logical state transfer and gross code automorphisms. We optimize transpiled algorithms scheduling towards the minimization of data movement.

We accomplish this for a specific architecture, modeling the transpiled algorithm comprising weight-constrained non-Cliffords and Cliffords as a directed acyclic graph (DAG). We prioritize executing operations whose qubits are entirely within the compute or memory regions with supported gate sets, allowing execution without movement. When qubit movement is unavoidable, the algorithm searches for which single qubit permutation between codes generates the largest set of executable gates. The appropriate gross-surface code is scheduled, and this process is iterated until the end of the DAG. This is seen in Figure \ref{fig:compilation_procedure}-(d),(e) and (f). The logical qubit structure of the gross code results in an almost constant-rate cost of logical qubit automorphisms. We find that the expected $X$ basis automorphisms is equal to almost always 2, similarly for the $Z$ basis with slightly higher variance. As such, we assume the access of any logical qubit in the gross code comes at the cost of $2$ X automorphisms, and $2$ Z automorphisms.

\subsection{Transpiling to HetEC}

To map an algorithm to our architecture, we must adhere to system constraints, particularly ensuring we do not generate Pauli product measurements of weight \( w \) exceeding the capacity of our surface code (see Figures~\ref{fig:arch_fig}(b) and \ref{fig:compilation_procedure}(c)). While optimizing non-Clifford operators can reduce gate counts, it increases operator sizes, placing higher demands on computational capacity. In a heterogeneous setting, additional constraints arise as we cannot exceed the computational capacity of the surface code area. Accordingly, we modify our transpilation rule
\textit{Commute a non-Clifford operator \( P' \) behind a Clifford operator \( P \) if \( PP' = -P'P \) and the number of non-trivial operators in \( iPP' \) does not exceed the number of available surface code tiles.}
Due to these constraints, Cliffords persist through transpilation because we cannot prune them behind a maximum-weight Pauli product operator~\cite{litinski2019game}. This contrasts with the homogeneous case, where all Cliffords can be pruned. Fortunately, the gross code supports the full set of Clifford operations, enabling a "compute-in-memory" approach for Cliffords. How much time is needed to implement an in-memory Clifford rotation on the gross code depends on the particular rotation. Native rotations are defined as those taking the least time, just two logical Pauli measurements or 14 syndrome cycles (see Table I). Other Clifford rotations must be synthesized as a sequence of native rotations \cite{cross2024linear} with an average of about 8 native rotations required. It is a complex transpilation problem, which we leave to future work, to route qubits so as to minimize the time of in-memory operations in general. Fortunately, there are four logical qubits that support native single qubit rotations $X(\pi/4)$ and $Z(\pi/4)$ and three logical qubit pairs that support native two-qubit Clifford rotations. Therefore, assuming it is possible to use precisely these qubits for in-memory operations, we can simplify the calculation by quantifying each in-memory Clifford rotation as 14 cycles. We also note that a relatively small fraction of gates are done in-memory for the example circuits we study, for example, around 2\% for the Adder and 3\% for the QFT.

Moreover, in a heterogeneous gate set, managing data movement between the surface code and gross code blocks becomes critical. Non-Clifford computations require all involved qubits to reside within the surface code patches. Therefore, data movement between codes must be carefully scheduled to avoid unnecessary overhead, extending the transpilation and compilation procedure to include:

\begin{itemize}
    \item \textbf{Inter Code Data Movement Calls}: Transferring qubits between gross codes and surface code tiles. Data movement should be minimized.
    \item \textbf{Internal Data Management}: Managing logical qubit positioning within the gross codes to facilitate efficient computation. This manifests as code automorphisms in the code, acting similar to that of a classical shift register. 
\end{itemize}

Because data movement between codes can be slower than executing non-Clifford and Clifford gates in the surface code, we must account for asynchronous operations within HetEC and schedule them accordingly.  This consideration is highlighted in Figures~\ref{fig:compilation_procedure}(a), (c), (d), and (e).

\subsection{HetEC Performance: Logical Clock Speeds}

Operating heterogeneously introduces challenges in aligning logical clock speeds, as each operation requires a different number of physical operations and syndrome measurements, with each operation requiring a different measurement cycle to be fault tolerant. For example, measuring a logical Pauli operator on the surface code fault tolerantly might require $d$ rounds of measurement, to a non-Clifford layer requiring anywhere from $d$ to $3d$ rounds of measurement based on the outcome of each fault tolerant logical measurement \cite{litinski2019game,bravyi2024high}. Scheduling operations within each code, and between codes, requires defining a base logical clock speed that the codes are both defined in. We define the global logical clock speed as one measurement cycle, under the assumption that measurements are the costliest operation in terms of time. This ties all operations within the architecture to the same clock speed, and makes synchronization of logical operations possible. We use these logical clock speeds as our unit of time, and to schedule logical circuits according to the proposed gate times.

\section{Related Work}

Research on designing heterogeneous error correcting architectures has begun seeing significant attention, as qLDPC codes are further investigated and improved upon. Examples of this includes the heterogeneous design of a qLDPC Surface Code architecture in a neutral atom architecture \cite{xu2024constant}, the integration of multiple codes for a hybrid memory-compute architecture \cite{thaker2006quantum}, or the design of concatenated qLDPC codes built on top of surface codes \cite{pattison2023hierarchical}. \cite{bravyi2024high} reports thresholds and specific qLDPC code design, and specifically the gross code, the qLDPC code we focus on in our research. \cite{thaker2006quantum} investigates a similar paradigm of a hybrid code architecture, though was before examples of high rate codes had been demonstrated, and hence is more focused on computing across codes of varying thresholds. Investigations into characterizing concatenated code threshold and performance \cite{rahn2002exact,jochym2014using,knill1996concatenated}, highlight another potential use case of heterogeneous error correction. However, we are focused on the discovery of how to run and transpile to these hybrid architectures, considering their constrained surface code size, asynchronous logical clock cycles, and logical qubit routing and placement within the gross code. To our knowledge, we are the first work to investigate the operation and end-to-end cost of running algorithms in the heterogeneous setting, and in doing so, discover and tackle novel computational challenges. 

\begin{table*}[t]
    \centering
    \renewcommand{\arraystretch}{1} 
    
    \begin{tabular}{c||c|c|c|c|c|c|c|c|c}
     \hline
     \hline
     Algorithm  & Qubits & \makecell{Phys. \\ Qubits \\ (All SC)} & \makecell{Phys. \\ Qubits \\ (2 SC)} & \makecell{Change \\  in \\ Phys. \\ Qubits} & \makecell{Logical Error \\ Rate \\ ($p=1e^{-3}$) \\ (2 SC, All SC)} & \makecell{Change \\ in \\ Error \\ Rate} & \makecell{Clock \\ Cycles \\ (All SC)} & \makecell{Clock \\ Cycles \\ (2 SC)} & \makecell{Increase in \\ Logical \\ Clock Cycles} \\
     \hline
     Adder & 18 & 8243 & 1355 & 6.08x & 1.38e-3, 7.29e-4 & 1.89x & 1596 & 4060 & 2.54x \\
     \hline
     Adder & 28 & 13369 & 1746 & 7.66x & 2.48e-3, 9.92e-4 & 2.50x & 1708 & 5859 & 3.43x \\
     \hline
     Adder & 64 & 29610 & 3780 & 7.83x & 3.98e-3, 2.38e-3 & 1.67x & 4158 & 6293 & 1.51x \\
     \hline
     Ising Model & 26 & 12078 & 1746 & 6.92x & 2.48e-3, 1.71e-3 & 1.45x & 2044 & 6727 & 3.29x \\
     \hline
     Ising Model & 42 & 19357 & 2137 & 9.06x & 4.16e-3, 2.79e-3 & 1.49x & 2464 & 4172 & 1.69x \\ 
     \hline
     Ising Model & 98 & 45420 & 4092 & 11.10x & 2.09e-3, 5.11e-4 & 4.09x & 2828 & 6034 & 2.13x \\
     \hline 
     QFT & 18 & 8243 & 1355 & 6.08x & 2.38e-3, 2.86e-3 & 0.83x & 16044 & 50239 & 3.13x \\
     \hline
    \end{tabular}
    \caption{Transpilation results over 2-surface code heterogeneous and solely surface code homogeneous architectures. We compare the space (physical qubit count), time (clock cycles), and error rate tradeoffs. 2 SC refers to a 2-surface code architecture, and all SC a homogeneous surface code architecture. }
    \label{tab:results}
\end{table*}

\section{Evaluation}

We focus on three key factors when executing a fault tolerant circuit: the logical error rate, physical qubit count, and execution depth. Operating heterogeneously is expected to incur increased time costs due to data movement overhead between codes. Data movement between codes is expected to increase logical error rate of circuits, though increasing pauli weights of non-cliffords does result in increased logical errors. However, heterogeneous architectures are expected to have substantial physical qubit savings. With near-term challenges being simply engineering a system of sufficient qubit quality and size, we want to investigate whether, given a specific physical error rate and physical qubit count, we should operate heterogeneously, with higher distance surface codes coupled to the gross code or homogeneously, comprising solely relatively lower distance surface codes. 

\subsection{Experimental Settings}

We explore the architectural trade offs in the following settings:

\textbf{Firstly,} we explore operating the physical qubit overheads of operating the gross code with cliffords support, hence having a capacity of 11 logical qubits. We explore how operating in this heterogeneous setting 
affects logical clock cycles, error rates, and physical qubit overhead.

\textbf{Secondly,} we explore how physical error rates affect logical error contributions over varying number of surface code modules.

\textbf{Finally,} we explore the physical and time cost overhead of attaining a specific logical error rate for the heterogeneous and homogeneous system. We tune surface code distance in the homogeneous setting to generate the target algorithmic logical error rate, and compare this system's overhead to the heterogenous setting. 

\begin{figure*}
    \vspace{1em} 
    \begin{subfigure}[t]{1.0\textwidth}
        \centering
        \includegraphics[width=\linewidth]{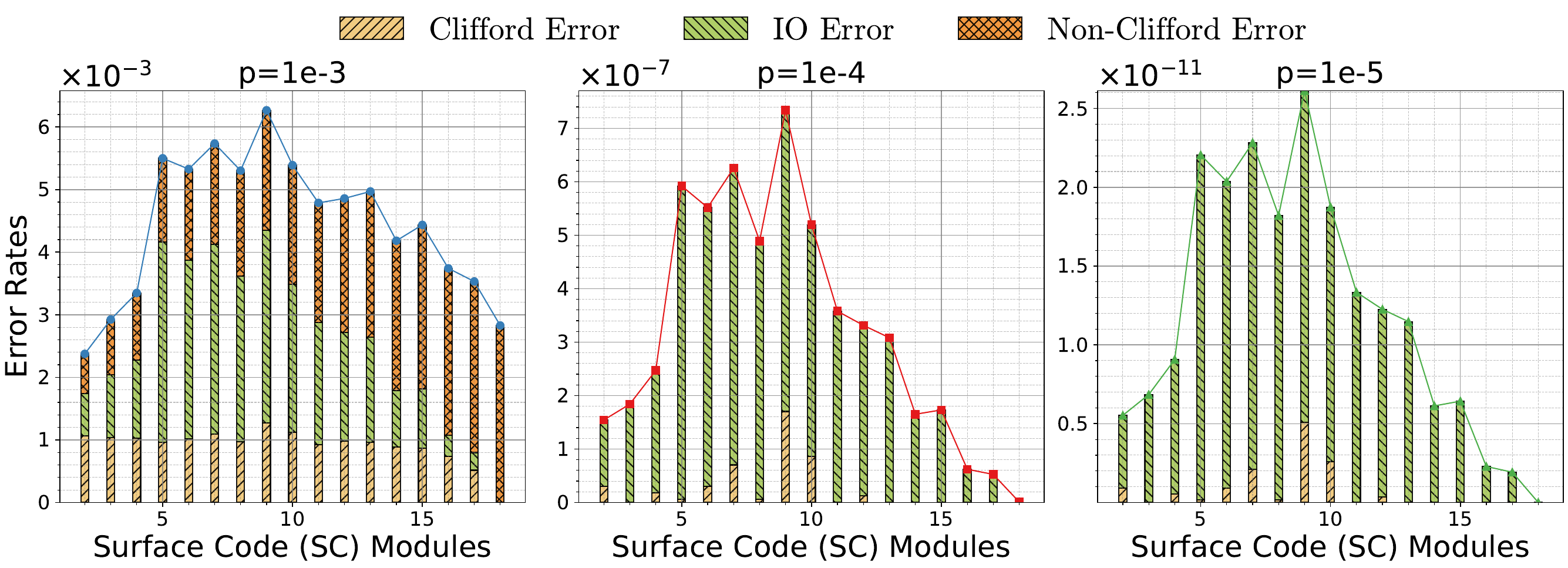}
    \end{subfigure}
\caption{Logical Error Breakdown of transpiled Quantum Fourier Transform over 18 qubits across multiple physical error models}
\label{fig:logical_error_qft}
\end{figure*}

\begin{figure*}
    \vspace{1em} 
    \begin{subfigure}[t]{1.0\textwidth}
        \centering
        \includegraphics[width=\linewidth]{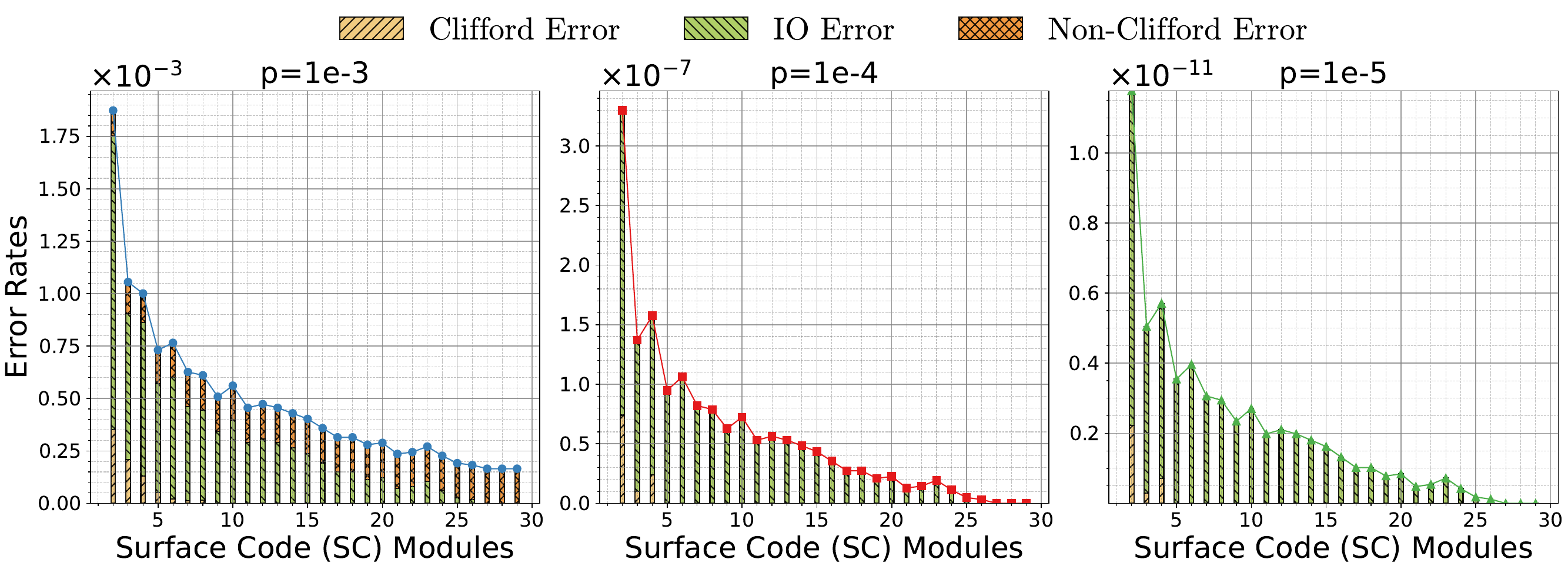}
    \end{subfigure}
    \caption{Logical Error Breakdown of a transpiled adder over 28 qubits across multiple physical error models}
    \label{fig:logical_error_adder}

\end{figure*}

We explore the transpilation of the Quantum Adder, a single layer of Ising Model simulation (representing a layer of digital quantum simulation), and the Quantum Fourier Transform to our architecture. QASM is sourced from \cite{li2023qasmbench}, and \cite{ross2014optimal} is used to transpile $R_z$ gates into Clifford + T gates. Each of these algorithms serves as a backbone for future algorithms, such as Shor's algorithm and digital quantum simulation. We evaluate their transpiled circuit structure with respect to Clifford gates, non-Clifford gates, and Input/Output (IO) count for the gross code block.

It is important to note that our gross code memory transpiler is not optimal. This is a challenging problem that suffers from combinatorial explosion when determining how to map logical qubits. As a result, we use a greedy approach, which introduces a small degree of randomness into our results. This explains the minor stochasticity observed in our plots.

\subsubsection{Circuit Fidelity Estimator}

Based on the instructions defined above, we estimate circuit fidelity by summing errors from each operation under each individual system setting. This is achieved by a two-step error rate simulation protocol, where we simulate the logical error rate of each logical operator, then substitute the original circuit with the logical error rates achieved from the prior simulation. The estimator could be further improved by not only simulating the independant local logical error rates for every instruction, but also to extend to a full error model capturing further dynamics of the heterogeneous system.

We approximate logical error rates of non-clifford operators as the weight of the pauli operator multiplied by $2$ times the non clifford error rate. This approach will assume that there is no ancilla routing overhead in the surface code tiling, and hence this forms the lower bound of this error rate. Longer ancilla routing is expected to incur additional overhead, and is more challenging for homogeneous systems as pauli operator weights increase in size.

\subsection{HetEC Evaluation}

\subsubsection{Comparison of Algorithm Error Rates}

\begin{figure}
    \centering
    \includegraphics[width=1.0\linewidth]{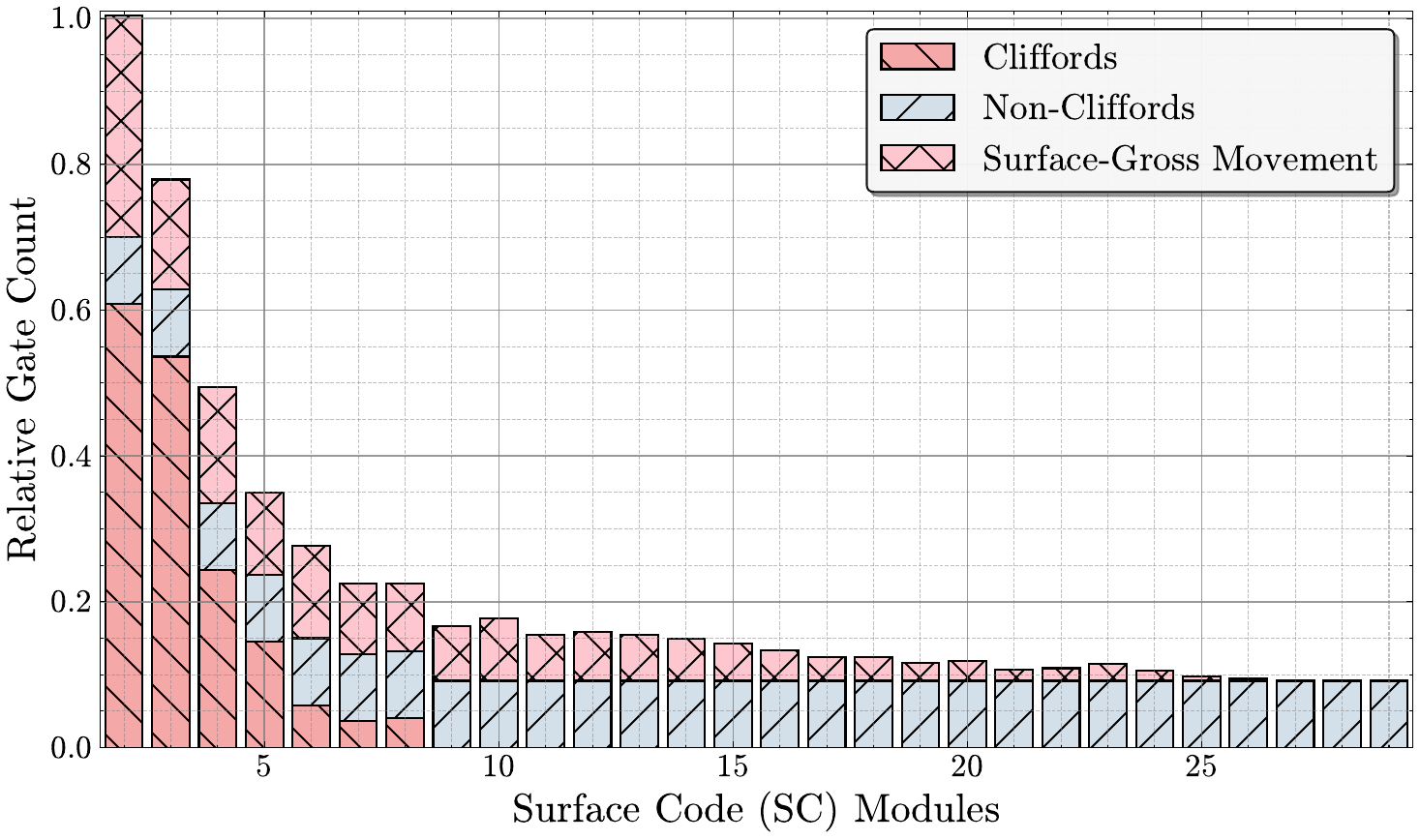}
    \caption{Gate distribution for the Adder 28, decomposed within the HetEC framework across various surface code modules. The surface-gross movement represents the input/output (IO) operations between surface codes and gross codes.}
    \label{fig:adder_gate_dist}
\end{figure}

When comparing a heterogeneous setup with two surface code patches to a homogeneous all-surface code system, we observe a trade-off in the Adder and Ising model circuits: lower physical qubit overhead comes at the cost of higher error rates. This increase in errors primarily arises from two factors: data movement overhead (labeled as IO Error) and additional Clifford gates. For example, in the 28-qubit Adder (see Figures~\ref{fig:adder_gate_dist} and \ref{fig:logical_error_adder}), data movement accounts for $56.4\%$ and Clifford gates for $11.8\%$ of the total error at $p = 1 \times 10^{-3}$. As the number of surface code patches increases, errors from non-Clifford gates become dominant, raising the non-Clifford error rate by up to $35.9\%$ and requiring $243\%$ more logical clock cycles compared to all-surface code systems.

These trends hold for the Adder and Ising models but differ for the Quantum Fourier Transform (QFT). In the QFT (Figure~\ref{fig:logical_error_qft}), optimal performance occurs either at low or high surface code counts. At low counts, non-Clifford errors are minimized due to limited operator weights (see Fig \ref{fig:compilation_procedure}-(c)), despite persistent IO and Clifford errors. At high counts, IO and Clifford errors decrease, but non-Clifford errors increase due to higher operator weights. Mid-range surface code counts result in the highest error rates, driven by significant IO costs, persistent Clifford errors, and high-weight non-Clifford operators. 

The QFT requires all qubits to interact with a single target qubit, leading to high-weight Pauli products. For instance, the Adder's average non-Clifford operator weight is $2.78$, whereas the QFT's is $6.95$ without operator weight constraints. This explains the bell-shaped error curve observed in Figure~\ref{fig:logical_error_qft}. At mid-range surface code counts (4--16), frequent data movement is required due to many high-weight non-cliffords, with many of the non-cliffords requiring data movement to abide by the surface code capacity constraint. In contrast, the Adder and Ising models, with relatively lower entanglement complexity, allow longer sequences of non-Clifford gates with fewer movement-induced errors.

Operating heterogeneously significantly reduces physical overhead--by an average factor of $7.81$--but increases the logical error rate by an average factor of $1.99$ and the logical clock cycles by $2.53$. Importantly, although the logical error rate increases, this is less problematic than in NISQ computation as errors can be exponentially suppressed by linearly increasing code distances. For circuits with high-weight non-Clifford gates, determining the optimal way to operate heterogeneously is more complex. Application-specific simulations are needed to assess whether the error increases from large non-Clifford operators outweigh the overhead benefits of heterogeneous operation.

\subsubsection{Comparison of Physical Error Models}

We explore physical error models at $p=1 \times 10^{-3}$, $p=1 \times 10^{-4}$, and $p=1 \times 10^{-5}$. While surface code error rates have been extensively studied, the error rates of gross code operations have been investigated less and are important for understanding HetEC. At $p=1 \times 10^{-3}$, all error channels are prevalent. In Figures \ref{fig:logical_error_adder} and \ref{fig:logical_error_qft}, the logical error rates across different channels are of similar magnitude. For example, in the 98-qubit Ising model, we observe $11.3\%$ IO error, $14.8\%$ Clifford error, and $73.9\%$ non-Clifford error. 

When the physical error rate decreases, e.g., $p=1 \times 10^{-4}$ and $p=1 \times 10^{-5}$, the IO errors dominate. This is due to the current interpolated logical error rates of state teleportation being much higher at these error rates, which alters the contribution of each error channel. If  IO errors are suppressed at a much lower rate compared to surface code operations, the logical errors arising from heterogeneous operations will increase accordingly. Gross codes, in particular, have not been thoroughly explored in terms of how errors are suppressed at lower physical error rates, so we place emphasis on the fact that $p=1 \times 10^{-4}$ and $p=1 \times 10^{-5}$ motivate seeking improvements in the error suppression of IO operations at these lower physical error rates. 

These results highlight that operating heterogeneously does not provide the same benefits at all physical error levels. It is important to consider the extent to which each operation's errors are exponentially suppressed as a function of the physical error rate. If certain codes within the system are suppressed to a lesser degree than others, their associated error channels will become more prominent as the physical error rate decreases.

\subsubsection{Comparison of Physical Overhead for Specific Algorithm Logical Error Rates}

In the previous sections, we examined algorithmic logical error rates for a fixed surface code distance. However, suppressing logical error rates may not be the best metric for fault-tolerant systems. Instead, it may be more useful to evaluate which system performs better at achieving a target logical error rate for a specific algorithm. This approach is important because reducing the code distance of the surface code can lower physical qubit overhead while still achieving better logical error rates compared to heterogeneous operations. We analyze this by determining the minimum code distance, and from here its physical qubits overhead, at which the surface code meets the target logical error rate for the algorithm.

\begin{figure*}[htbp]
    \centering
    
    \begin{subfigure}[t]{0.33\textwidth}
        \centering
        \includegraphics[width=\linewidth]{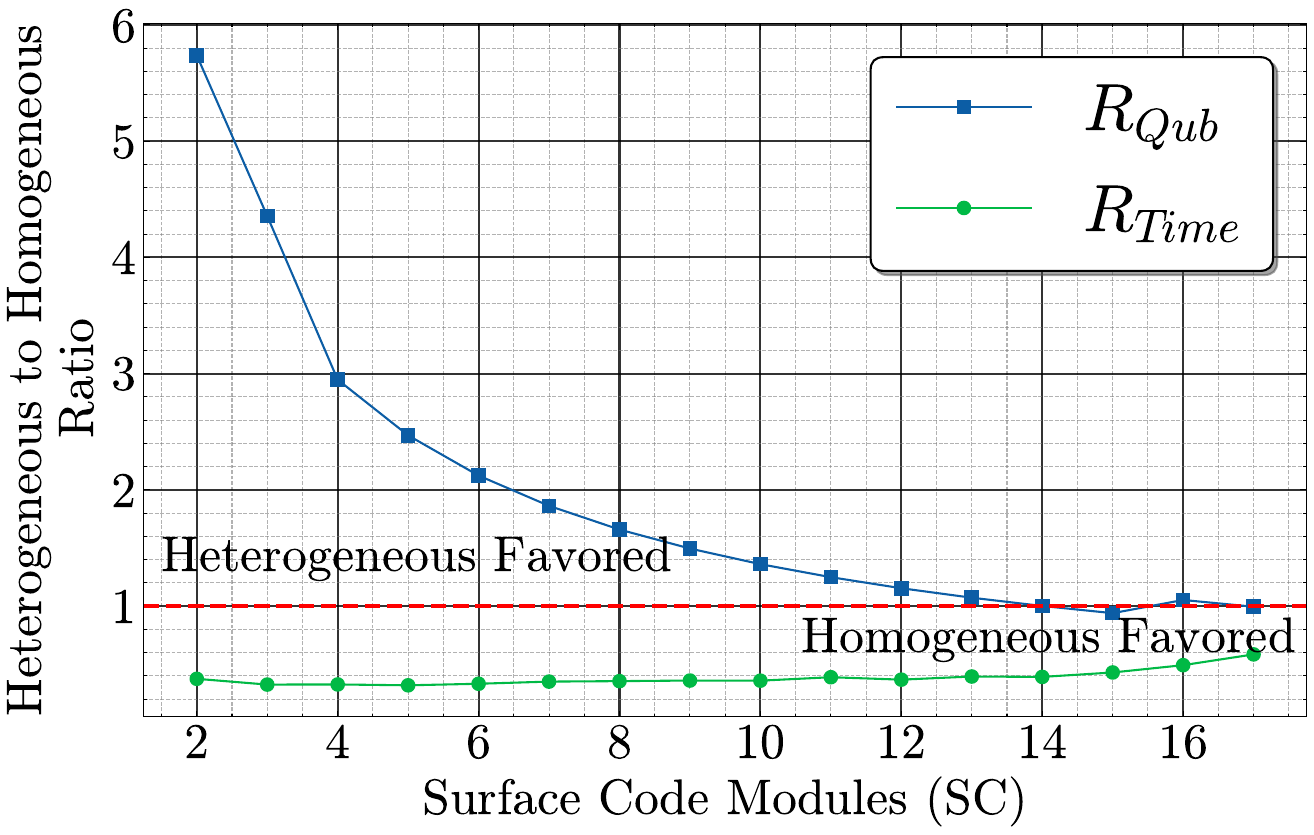}
    \caption{Quantum Fourier Transform}
        \label{fig:qubit-count}
    \end{subfigure}
    \hfill
    \begin{subfigure}[t]{0.33\textwidth}
        \centering
        \includegraphics[width=\linewidth]{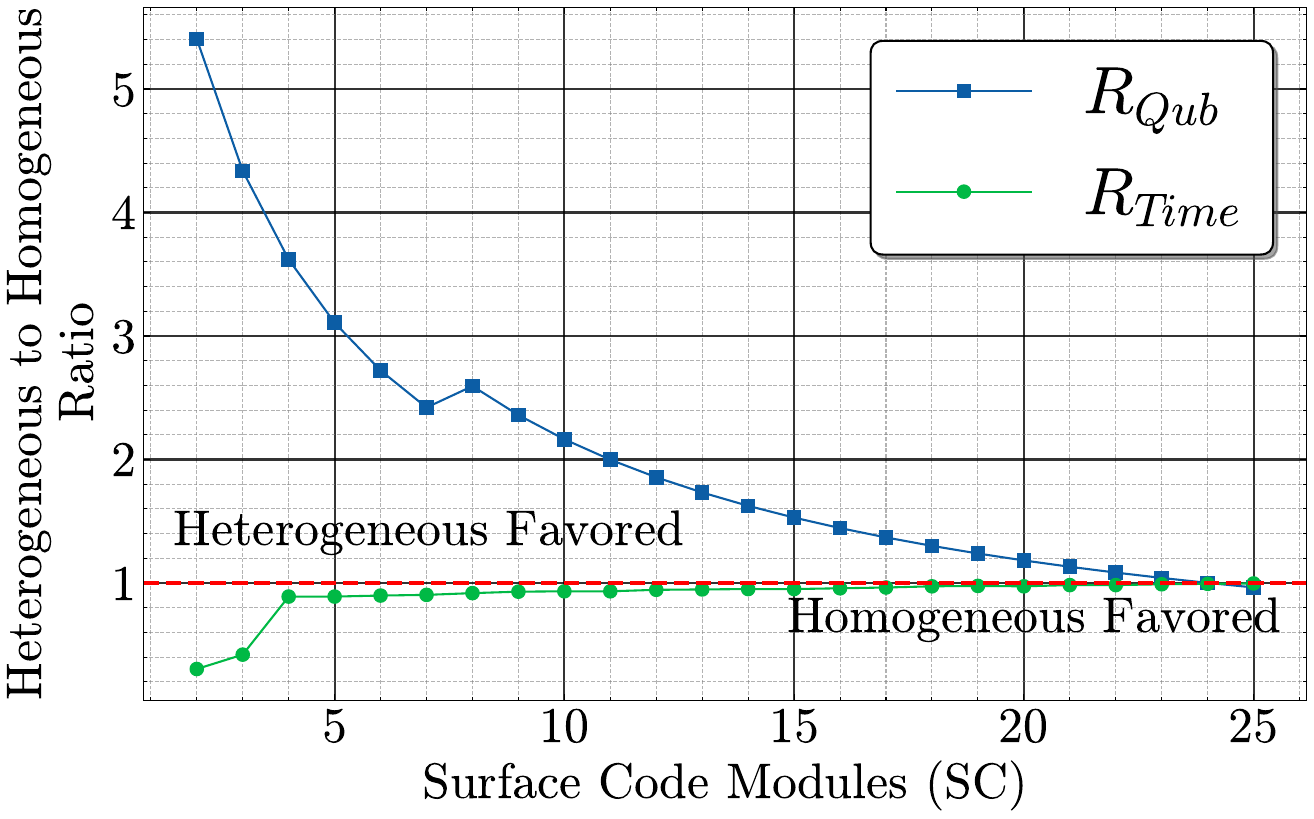}
        \caption{Ising Model Simulation}
        \label{fig:qft-io}
    \end{subfigure}
    \hfill
    \begin{subfigure}[t]{0.33\textwidth}
        \centering
        \includegraphics[width=\linewidth]{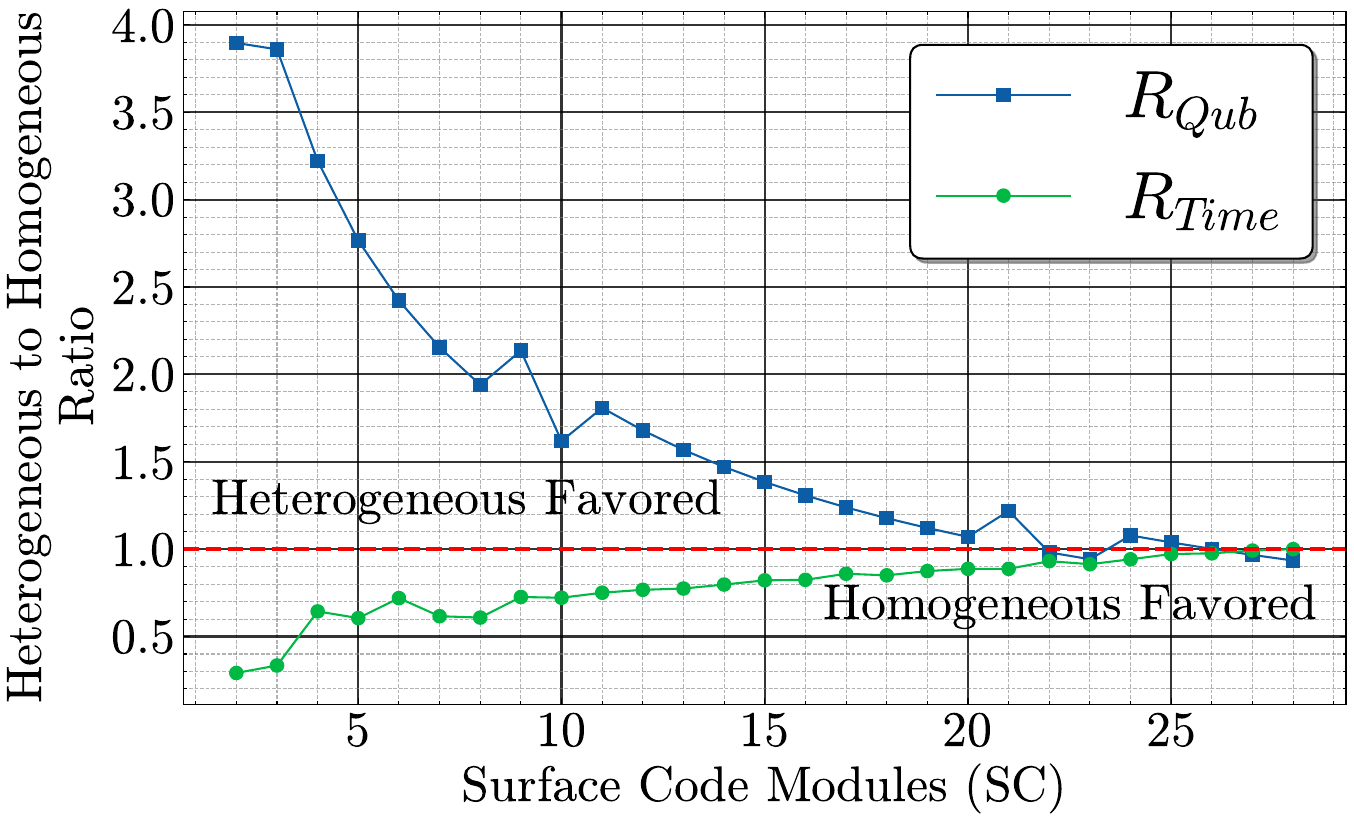}
        \caption{Adder}
        \label{fig:gate-count}
    \end{subfigure}
\caption{Tradeoff between heterogeneous and homogeneous operation for executing an algorithm to achieve a specific logical error rate, as discussed in Section 4.2.2. $R_{Qub}$ represents the ratio of physical qubits in the heterogeneous system to those in a surface code system with distance $d$ that achieves the same specific logical error rate $e$. $R_{Time}$ indicates the relative slowdown in heterogeneous operation, with $\frac{1}{R_{Time}}$ reflecting the corresponding increase in logical clock cycles.}
\label{fig:qubit_time_tradeoff}
\end{figure*}

We set the target algorithm's logical error rate $e$ to match the rate achieved by a 2-surface code heterogeneous system at a physical error rate of $p = 1 \times 10^{-3}$ (denoted as 2-SC, $p = 1 \times 10^{-3}$ in Table \ref{tab:results}). Next, we determine the code distance $d$ required for a pure surface code architecture to achieve the same error rate. This gives us the corresponding physical qubit overhead for the surface code at distance $d_{Surf}$. We then compute the relative physical qubit overhead, $q_{d_{Surf}}$, for this distance and compare it to the heterogeneous setting, which achieves the same error rate $e$ with overhead $q_{HetEC}$. The ratio of heterogeneous to homogeneous physical qubits is defined as $R_{Qub} = \frac{q_{HetEC}}{q_{d_{Surf}}}$. When $R_{Qub} > 1$, the heterogeneous system is favored, and when $R_{Qub} < 1$, the homogeneous system is favored.
We compute the relative time costs for each setting by calculating the logical clock cycles required for HetEC ($t_{HetEC}$) and a purely surface code architecture ($t_{SC}$). The ratio $R_{Time} = \frac{t_{SC}}{t_{HetEC}}$ represents the clock cycle ratio, indicating the relative slowdown when operating heterogeneously. Due to overhead from routing and Clifford operations in heterogeneous systems, these values will generally be $<1$, with $1$ corresponding to the homogeneous case. Values closer to $1$ indicate lower time costs in the heterogeneous setting.

These results are shown in Figure \ref{fig:qubit_time_tradeoff}. We observe up to a $6.42\times$ improvement in physical qubit overhead when operating heterogeneously, with a slowdown factor of $0.31$ for the 26-qubit Ising model. However, when increasing the surface code count from 2 to 4, the physical qubit overhead improvement drops from $6.42\times$ to $4.33\times$, but the slowdown factor improves to $0.91$. This represents a $32.6\%$ increase in qubit overhead for a $175.8\%$ speedup. In the case of the Adder, a $32.3\%$ reduction in physical qubit overhead results in a $90.6\%$ speedup. These findings demonstrate that optimizing a heterogeneous architecture requires careful consideration of the algorithm's structure.

With the current goal of fault-tolerant quantum computing being the successful execution of an algorithm at a sufficiently low error rate to extract meaningful information, these results are particularly promising. Operating heterogeneously offers significant reductions in physical qubit overhead when targeting specific logical error rates, even with reduced surface code distances. This is due to the fact that the logical error rates in Table~\ref{tab:results} remain of similar orders of magnitude across architectures; however, reducing surface code distance incurs an exponential increase in errors, as shown in Table~\ref{tab:surinst}. For example, in seeking the minimum distance \(d\) that achieves an algorithmic logical error rate \(e\), each step down in \(d\) leads to an exponential rise in error rates. Since logical error rates in heterogeneous setups are similarly scaled, further reductions in surface code distance would lead to exponentially higher errors. Given that surface codes yield worse than the gross code, this modest reduction in distance does not meaningfully lower the physical overhead compared to a gross code-based heterogeneous approach.

\section{Conclusion}

Operating error correcting architectures heterogeneously poses substantial promise in the traversal towards fault tolerance. HetEC provides an early evaluation of architectures for integrating heterogeneous error-correcting codes with high rate codes, hybrid pauli based computation and ancilla based state teleportation. We introduce the required transpilation techniques, compute memory logical qubit management schemes, and explicit simulation modeling and scheduling for these architectures. Our architecture, while focusing on combining the gross code with the surface code, is applicable to other code combinations. We selected these codes due to their respective strengths IBM's gross code being a qLDPC target with a favorable structure, and the surface code's ability to enable universal gate sets with magic states.

We demonstrate significant numerical trade-offs in physical qubit counts, logical clock cycles, and logical error rates, showing that heterogeneous architectures can execute algorithms to a specific precision with substantially lower physical qubit overheads—up to 6.42x at the same physical error rate. As the traversal toward fault tolerance progresses, these system-level optimizations will be as crucial as hardware advancements, highlighting the potential of heterogeneous architectures in minimizing resource demands for error correction.

\section*{Acknowledgements}

We would like to thank Mark Ritter and Margaret Martonosi for their insightful discussions and contributions to this work, and Patrick Rall for providing data regarding Clifford synthesis in the gross code. This material is based upon work supported by the U.S. Department of Energy, Office of Science, National Quantum Information Science Research Centers, Co-design Center for Quantum Advantage (C2QA) under contract number DESC0012704, (Basic Energy Sciences, PNNL FWP 76274). This work is in part funded by National Science Foundation (NSF) under award CCF-2338063. Yongshan Ding receives consulting fees from Quantum Circuits, Inc. This research used resources of the National Energy Research Scientific Computing Center (NERSC), a U.S. Department of Energy Office of Science User Facility located at Lawrence Berkeley National Laboratory, operated under Contract No. DE-AC02-05CH11231. This research used resources of the Oak Ridge Leadership Computing Facility, which is a DOE Office of Science User Facility supported under Contract DE-AC05-00OR22725. The Pacific Northwest National Laboratory is operated by Battelle for the U.S. Department of Energy under Contract DE-AC05-76RL01830.

\clearpage

\bibliographystyle{ACM-Reference-Format}
\bibliography{refs}

\end{document}